\documentstyle[epsf,12pt]{article}
\newcommand{\bq}{\begin{equation}}
\newcommand{\eq}{\end{equation}}
\newcommand{\ba}{\begin{eqnarray}}
\newcommand{\ea}{\end{eqnarray}}
\newcommand{\nobody}{\rule{0ex}{1ex}}

\begin{document}
\thispagestyle{empty}
\begin{flushright}
MPI-PhT/95-36\\ LMU-08/95\\ hep-ph/9505298 \\May 1995\vspace*{3cm}
\end{flushright}
\begin{center}
{\LARGE\bf Production of $B_c$ Mesons in Hadronic
Collisions\footnote
{Supported by the German Federal Ministry for Research
and Technology under contract No.~05~6MU93P.}}
\vspace{2cm}\\
Karol Ko\l odziej$^{\rm a,}$\footnote{
        On leave from the Institute of Physics, University of Silesia,
        PL--40007 Katowice, Poland},
Arnd~Leike$^{\rm a}$
        and
Reinhold~R\"uckl$^{\rm a,b}$\vspace{0.5cm}\\
$\nobody^{\rm a}${\small\it
Sektion Physik der Universit\"at M\"unchen,\\
 Theresienstr. 37, D--80333 M\"unchen, FRG}\\
$\nobody^{\rm b}${\small\it
Max-Planck-Institut f\"ur Physik, Werner-Heisenberg-Institut,\\
F\"ohringer Ring~6, D-80805 M\"unchen, FRG}
\vspace*{4cm}\\
{\bf Abstract}
\end{center}
{\small
We investigate hadroproduction of $B_c$ mesons from initial gluons
 using QCD perturbation theory and
nonrelativistic bound state approximations. Our results obtained by two
completely independent calculations are confronted with existing
results which contradict each other.
Moreover, we examine the
approximation based on heavy quark fragmentation
 and determine the range of validity of it.
Predictions for cross sections and differential distributions at
Tevatron and LHC energies are presented.
}
\vfill\newpage
%
\section{Introduction}
With the discovery of the top quark, it has become rather certain that
$B_c$ mesons
are the only flavored heavy quark resonances which exist in nature.
Although they have not yet been observed experimentally, the production
rates at the Tevatron
may be sizeable enough to
allow for experimental identification
of these interesting bound states. Clear evidence can be expected
from the LHC \cite{atlascms},
where it should also be possible to measure specific
production and decay properties. This will provide valuable tests
of important aspects of strong and weak interactions.

Because of the promising prospects, $B_c$ -- physics is attracting
increasing attention. In particular, the hadronic production of $B_c$
mesons has been studied by several groups \cite{cc}--\cite{ms}
in the framework of QCD perturbation theory treating
the $B_c$ couplings and wave functions
in a nonrelativistic approximation. There is general agreement
that in $pp$ and $p\bar{p}$ collisions
at high energies the gluon-gluon scattering process
$gg\rightarrow B_cb\bar c$
including higher order corrections is by far the dominant source
of $B_c$ mesons.
However, the quantitative predictions derived first in
refs. \cite{cc,s,bls} in lowest order contradict each other
in various respects. It is not always easy to compare the results,
because of the use of different, sometimes outdated gluon momentum
distributions and the lack of clear statements on the
choice of scales and bound state parameters. Yet, one can convince
oneself that differences in the numerical input alone cannot explain the
discrepancies. The problem is most obvious in refs. \cite{cc} and \cite{bls}
which disagree already on the level of the subprocess. The confusion
has also not been resolved in ref. \cite{ms} which appeared very recently
while we were writing up this paper.
The situation is rather unsatisfactory and needs clarification.

Another controversial issue is the description of $B_c$ hadroproduction
in terms of $b$-pair production in $gg$--fusion
followed by fragmentation of the $\bar{b}$-quark:
$gg \rightarrow b\bar{b},\ \bar b\rightarrow B_c b\bar c$.
The application of this approximation \cite{PP2} was motivated
by the success of the analogous approach
to $B_c$ production in $e^+e^-$ annihilation.
In fact, from the perturbative calculation of
$e^+e^- \rightarrow B_c b\bar c$ \cite{ee}
one can directly extract the relevant fragmentation function
$D_{\bar{b}}(z)$, $z$ being the momentum fraction of the $\bar b$-quark
transferred to the $B_c$ meson \cite{fragee, teup94}.
In hadronic processes, the hard scattering formalism involving convolutions
of parton cross sections with structure and fragmentation functions
is generally expected to work for high--$p_T$ production.
On the other hand, it has been shown
in refs. \cite{gaga, klr} that the fragmentation picture is not adequate
for $B_c$ production in photon-photon scattering, not even at large
$\gamma\gamma$ energies and the highest $B_c$ transverse momenta.
Since the two processes
$\gamma\gamma \rightarrow B_cb\bar c$ and $gg \rightarrow B_cb\bar c$
are very similar at least if one
disregards the gluon self--interactions and the color factors,
one may mistrust the general expectation expressed above.
While the previous investigations of this issue in refs. \cite{cc}
and \cite{bls} are not conclusive because of the discrepancies of these
calculations, ref. \cite{bls1} which also appeared just recently
 claims the failure of the fragmentation approximation. We will
clarify  this important question and show that the fragmentation
description indeed works to the expected accuracy.

In order to achieve these goals, we have performed two independent
calculations of the dominant  subprocess
$gg\rightarrow B_cb\bar c$ in $O(\alpha_s^4)$ and in the usual nonrelativistic
approximation. The result is compared numerically with the previous
calculations.
Where a comparison at the level of the subprocess was
 not possible, we have checked the convoluted cross sections for
the same parton distributions and input parameters
as used in the calculation under examination.
For the integrated cross sections, we get partial
agreement with ref. \cite{bls},
however for a different value of $m_b$
and only after including a factor 1/3. Although this missing factor
has been corrected in ref. \cite{bls1}, there are still discrepancies,
in particular, in $p_T$--distributions.
Furthermore, we clearly disagree with refs. \cite{cc,s,ms}.

For predictions on $B_c$
production  at the Tevatron and LHC  the $gg$
cross section is folded with the updated gluon distribution of
ref. \cite{mrs} in which  the HERA measurements are taken into account.
Moreover, we
study in detail the dependence on the scale chosen for $\alpha_s$ and the
structure functions.
Finally, we present distributions in transverse momentum and rapidity,
as well as integrated cross sections including $p_T$-- and $y$--cuts.

The paper is organized as follows. In chapter 2 we give
details of the calculation of the matrix element and the phase space
integration. There, we also specify the nominal values of all relevant
parameters used in our numerical studies. Chapter~3 is devoted to
 total cross
sections, and chapter~4 to differential distributions. A short summary
and conclusions are given in chapter 5.

%
\section{Calculation of the subprocess $gg\rightarrow B_c^{(*)}b\bar c$}
To  order  $\alpha_s^4$, the subprocess
$gg\rightarrow B_c^{(*)}b\bar c$ is described by
36 Feynman diagrams.
For the heavy $b$ and $c$ quarks, it is reasonable to neglect
 the relative momentum
of the quark constituents and their binding energy relative to  their masses.
In this nonrelativistic limit, the constituents are on mass shell and move
together with equal velocity. This implies the following relations for
the masses and momenta of the  $c$-quark,
$\bar b$-quark, and the $B_c^{(*)}$ bound state:
\begin{equation}
\label{bound}
M(B_c)=M(B_c^*)=M=m_c+m_b,\ \ \ p_c = \frac{m_c}{M} p,
\ \ \ p_{\bar{b}} = \frac{m_b}{M} p.
\end{equation}
Furthermore, the amplitudes for the production of
$S$--wave states are obtained from the hard
 scattering amplitudes for $gg\rightarrow b\bar b c\bar c$ by applying the
projection operator \cite{GKPR,klr}
\begin{equation}
\label{nonrela}
v(p_{\bar b})\bar u(p_c) = \frac{f_{B_c^{(*)}}}{\sqrt{48}}(p\!\!/ - M)
                           \Pi_{SS_Z}\; ,
\end{equation}
where $v(p_{\bar b})$ and $\bar u(p_c)$ denote the respective quark spinors and
$\Pi_{SS_Z}$ is a spin projector. For a
pseudoscalar state one has $\Pi_{00}=\gamma_5$, while for a
 vector, $\Pi_{1S_Z}=\rlap/\epsilon$.
{}From now on, we differentiate between the pseudoscalar ground state $B_c$
and the vector ground state $B_c^*$.
The decay constants $f_{B_c^{(*)}}$ appearing in eq. (\ref{nonrela})
are defined by the matrix elements
\bq
<0|\bar b\gamma_\mu\gamma_5 c|B_c(p)>\  =\  if_{B_c}p_\mu,\ \ \
<0|\bar b\gamma_\mu c|B_c^*(p)>\  =\  Mf_{B_c^*}\epsilon_\mu,
\eq
and are related to the bound state wave function at the origin by
\bq
f_{B_c} = f_{B_c^{*}} = \sqrt{\frac{12}{M}}|\Psi(0)|.
\eq
As in ref. \cite{klr}, the color structure is not
accounted for in eq.~(\ref{nonrela}).
The color factors are computed separately when squaring the amplitude.

In order to have an internal cross-check, we have calculated
 the matrix element squared by two independent methods.
One calculation is based on the usual trace technique, while the other
 calculation makes use of helicity amplitudes similarly as in ref. \cite{klr}.
The latter method is a generalization of the technique
described in detail in ref. \cite{KZ} for bosonic final states in
$e^+e^-$ scattering.
The results of these two calculations are compared numerically for
different sets of particle momenta and found to agree to seven digits.
Furthermore, it has been checked that the matrix element
vanishes when
the polarization vector of anyone of the initial gluons is substituted
by the momentum vector of this gluon.

The phase space integration is also performed independently using two
different routines.
In a first step, the integration routines are checked for the pure
3-particle phase space and found to agree numerically
with the corresponding analytical
 formulae to four digits for all distributions considered.
Before integrating  the matrix element squared with the help of
 the Monte Carlo
routine {\tt VEGAS}~\cite{vegas}, the strongest
peaks of it were smoothed out by introducing new integration variables.
After the integration all differential distributions and cross
sections of the two independent calculations agree within the Monte Carlo
errors.

In addition, we have calculated cross sections and distributions
for $b\bar b(c\bar c)$
production and subsequent fragmentation of the
$\bar b(c)$--quark, that is for
$gg\rightarrow b\bar b,\ \bar b\rightarrow B_c\bar c$ and
$gg\rightarrow c\bar c,\ c\rightarrow B_cb$, respectively.
 The relevant fragmentation functions $D_{\bar{b}}(z)$ and $D_c(z)$
are known from perturbation theory \cite{fragee}.
In $e^+e^-\rightarrow B_c\bar c b$, they
provide a perfect approximation for the energy distribution
${\rm d}\sigma/{\rm d}z$
with an error of the order of $M^2/s,\ \sqrt{s}$ being the c.m. energy
\cite{fragee,teup94}.
In contrast, in $\gamma\gamma\rightarrow B_c\bar c b$ above threshold, the
 fragmentation mechanisms are dominated by quark recombination processes.
This can be shown in a gauge invariant way without actually factorizing
the process in heavy quark production and fragmentation \cite{klr}.
Because of the similarity of the underlying
processes, one anticipates recombination
dominance in  $gg\rightarrow B_c\bar c b$ at least at low $p_T$.
Concerning the fragmentation
mechanism, we  stress that the factorized description
${\rm d}\hat\sigma(gg\rightarrow b\bar b)\otimes D_{\bar b}(z)$
is not justified at
small $p_T$ and close to threshold, where the quark  masses are important.
This caveat is ignored in ref. \cite{bls1}.
In order to find out where and to what extent this approximation
can be trusted, we have also assumed various  relations between daughter and
parent momenta:
\bq
\label{prescription}
p_T = z\sqrt{\hat s/4 -\mu^2}\sin\theta_b,\ \ p_L=p_T\cot\theta_b,\ \
\mu=M\ ({\rm I}),\ \ m_b\ ({\rm II}),\ \ 0\ ({\rm III}),
\eq
where $\sqrt{\hat s}$ is the gluon--gluon c.m. energy.
The case I respects the physical phase space boundaries, II and III
are assumed in refs. \cite{PP2} and \cite{bls1}, respectively.
These prescriptions are equivalent to the accuracy of the fragmentation
approach.

To obtain cross sections for $pp$ or $p\bar{p}$ collisions,
the gluon--gluon cross
section $\hat\sigma$ has to be folded with the gluon structure functions
$g(x,Q^2)$ of the (anti)proton:
\bq
\label{fold}
\sigma(s)=\int_0^1{\rm d}x_1\int_0^1{\rm d}x_2\,g(x_1,Q^2)g(x_2,Q^2)
\theta(x_1x_2s-4M^2)\hat{\sigma}(sx_1x_2,Q^2).
\eq
Analogous convolution formulae hold for differential distributions.
For numerical illustrations we use the following input
\bq
f_{B_c}=f_{B_c^*}=0.4\,{\rm GeV},\ \ \ m_b=4.8\,{\rm GeV},\ \ \ m_c=1.5\,
{\rm GeV}
\label{input}
\eq
together with the {\tt MRS(A')} gluon distribution \cite{mrs}.
The strong coupling constant $\alpha_s(Q^2)$ entering $\hat\sigma$, is
taken in leading logarithmic approximation:
\begin{eqnarray}
\label{alphas}
\alpha_s(Q^2)= { \alpha_s(m_Z^2) \over {1+ {{33-2 n_f}\over {12 \pi}}
\alpha_s(m_Z^2) \ln ({Q^2\over m_Z^2})}}
\end{eqnarray}
with $n_f=5$ and  $\alpha_s(m_Z^2)=0.113$.
This is the normalization of the running coupling in
 the {\tt MRS(A')} parton distributions.
As is usually done in such calculations, we identify the scale of
 $\alpha_s$ with the evolution scale of  the structure functions.
\ \vspace{1cm}\\
\begin{minipage}[t]{7.8cm} {
\begin{center}
\hspace{-1.7cm}
\mbox{
\epsfysize=7.0cm
\epsffile[0 0 500 500]{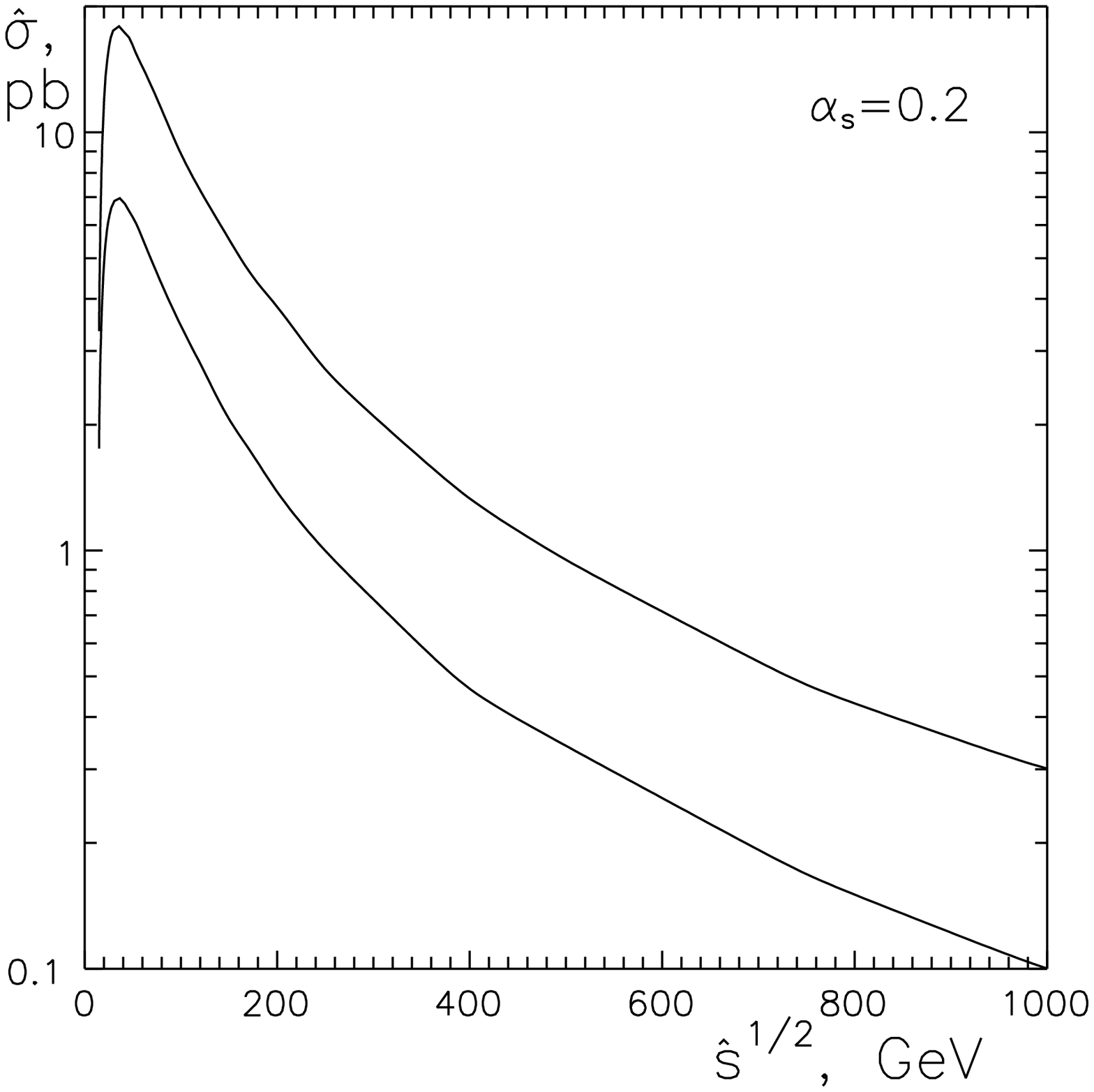}
}
\end{center}
\vspace*{-0.5cm}
\noindent
{\small\bf Fig.~1. }{\small\it
Integrated cross sections for $gg \rightarrow B_c^{(*)} b \bar{c}$
versus the c.m. energy.
The lower (upper) curve corresponds to $B_c$ ($B_c^*$) production.
}
}\end{minipage}
\hspace{0.5cm}
\begin{minipage}[t]{7.8cm} {
\begin{center}
\hspace{-1.7cm}
\mbox{
\epsfysize=7.0cm
\epsffile[0 0 500 500]{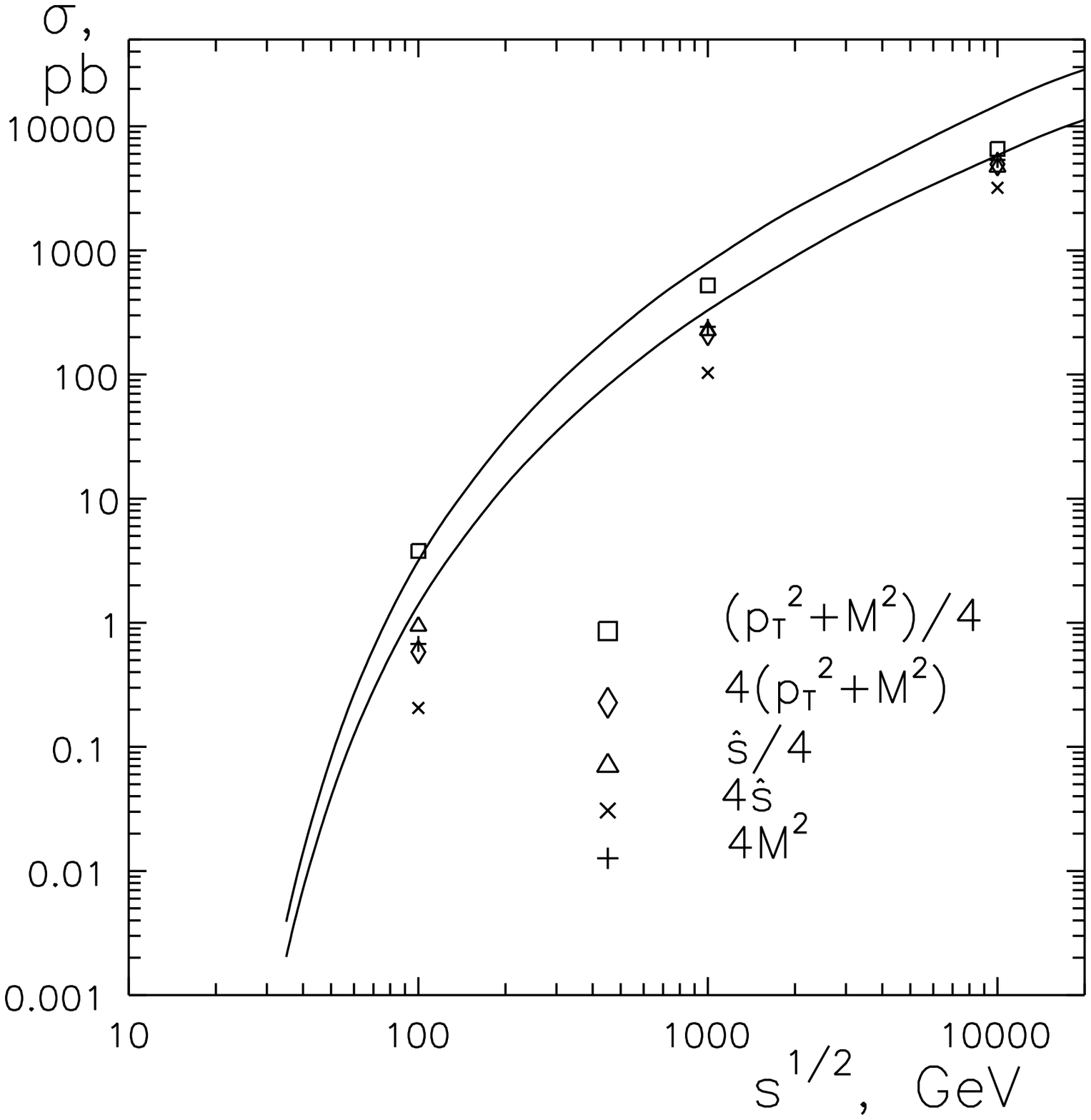}
}
\end{center}
\vspace*{-0.5cm}
\noindent
{\small\bf Fig.~2. }{\small\it
Total cross sections for $pp(p\bar p) \rightarrow B_c^{(*)} b \bar{c} + X$
versus the c.m. energy.
The lower (upper) curve corresponds to $B_c$ ($B_c^*$) production calculated
with the scale $Q^2=p_T^2+M^2$.
The symbols refer to other choices.
}
}\end{minipage}
\vspace*{0.5cm}
%
\section{Total cross sections}
First we present the parton cross sections $\hat\sigma(\hat s)$
for the subprocess $gg\rightarrow B_c^{(*)}b\bar c$, and
the convoluted production cross sections for
$pp({\rm\ and\ }p\bar{p})\rightarrow B_c^{(*)}b\bar c + X$.
Then we confront our results with results  in
the published and unpublished  literature \cite{cc,s,bls,ms}.

Figures~1 and 2 illustrate our main findings. The partonic
cross sections are shown in Fig.~1 for a fixed value of $\alpha_s$ in order to
 facilitate comparison with other calculations.
The production threshold is located at $\sqrt{\hat s}=2(m_b+m_c)=12.6\,$GeV.
At sufficiently high energies, the cross sections fall roughly like
$1/\hat s$. Moreover, the $B_c^*$ cross section is approximately three times
as large as the $B_c$ cross section as expected from spin counting.

The total hadronic cross sections for $B_c$ and $B_c^*$ production are
plotted in Fig.~2.
One can see the typical rise of  $\sigma$
 with energy reflecting the rise
of the gluon density as $x$ approaches $x_{min}=4(m_b+m_c)^2/s$, and
the peaking of $\hat\sigma$ near threshold.
A similar behaviour is observed for $B_c^{(*)}$ production
in the scattering of two bremsstrahlung photons radiated
from $e^\pm$--beams \cite{klr}.
The scale dependence is also exhibited in Fig.~2
for a variety of choices including the fixed scale $Q^2=4M^2$.
 At relatively low energies, the
predictions vary by more than one order of magnitude. At higher energies,
however, the results become more stable changing only within a factor two.
It is interesting to note in this respect that the differences
resulting from convolution with the gluon densities {\tt MRS(A')} of ref.
\cite{mrs} and {\tt CTEQ2} of ref. \cite{cteq} are very small.
In fact, they would be invisible in Fig.~2 .
In addition to the scale
uncertainty which can only be reduced by calculating the
next--to--leading corrections, a formidable task indeed, one has
uncertainties connected with the decay constants $f_{B_c^{(*)}}$ and the
effective quark masses $m_b$ and $m_c$. These amount at least
to another factor of two.

Next, we confront our results with the numerical results of refs.
 \cite{cc,s,bls,ms}.
The comparison is easiest and clearest for the parton cross section
$\hat{\sigma}$, where one does not have to pay attention
to different parametrizations of the gluon distribution and
for a constant value of $\alpha_s$.
Needless to say, we have always adopted the same values
for $m_b,\ m_c$ and $f_{B_c^{(*)}}$ as the reference calculation.

\begin{table}
\begin{center}
\begin{tabular}{|l|c|c|c|c|}
\hline
                                &  20 GeV  &  40 GeV  & 100 GeV  &  1 TeV   \\
\hline
\hline
$\hat{\sigma}_{B_c}$            &  8.27(2) & 12.7(1)  & 6.67(4)  & 0.208(2) \\
\hline
$\hat{\sigma}_{B_c}$, \cite{bls}&  29.3(1) &   40(1)  &  21(1)   &  0.27(4) \\
\hline
\hline
$\hat{\sigma}_{B_c^*}$       &   20.3(1)   &   33.1(2) &  17.4(1)&  0.61(1)  \\
\hline
$\hat{\sigma}_{B_c^*}$, \cite{bls}& 71.6(3) &  105(2)  &  56(2)    & 0.6(1) \\
\hline
\end{tabular}
\caption{Comparison of total cross
         sections in pb for $gg \rightarrow B_c^{(*)} b \bar{c}$ with the
         corresponding results of  ref. [4]. The input
         parameters are $m_b=5.1$ GeV, $m_c=1.5$ GeV, $\alpha_s=0.2$ and
         $f_{B_c}=f_{B_c^*}=0.57$ GeV. The number in parenthesis shows
         the Monte Carlo uncertainty in the last digit.}
\label{blstab}
\end{center}
\end{table}

 Table~\ref{blstab} summarizes the comparison with ref. \cite{bls}.
As can be seen, we clearly disagree with  this calculation.
Amazingly, if one uses instead of the quark masses quoted the masses given in
 eq.~(\ref{input}) one achieves agreement at
$\sqrt{\hat{s}} =20, 40$ and $100$\,GeV
within the Monte Carlo errors, up to a color factor 1/3
 which has apparently
been missed. In the recent repetition  \cite{bls1} of their calculation,
the authors account for the missing color factor.
However, we do not see an explanation for
the remaining discrepancy at $\sqrt{\hat s} = 1\,$TeV
which exceeds the MC uncertainty by far.
We also observe a clear discrepancy with ref. \cite{ms} as demonstrated in
Table~2. The fact that these authors claim rough
agreement with ref. \cite{bls} speaks against their results
because of the error in the  color factor pointed out above.

Furthermore, results on the $gg$--cross section can be found in
 the second paper of ref. \cite{cc}. Unfortunately, since
 the choice of the QCD--scale $\Lambda$ and of the momentum scale $Q^2$
is left unclear, a direct
 comparison is difficult. In Table~\ref{cctab},
we try two hypotheses: fixed $\alpha_s$ and $\alpha_s(\epsilon\hat s)$,
where the numerical factor $\epsilon$ can be chosen arbitrarily. As one can
see, in none of the two cases we are able to achieve agreement.
\begin{table}
\begin{center}
\begin{tabular}{|l|c|c|c|c|c|c|}
\hline
              & 20 GeV  & 30 GeV  & 40 GeV  & 60 GeV  & 80 GeV  & 100 GeV \\
\hline
\hline
$\hat{\sigma}_{B_c}$
              & 6.86(2) & 9.71(4) & 9.74(5) & 7.93(5) & 6.23(5) & 5.01(4) \\
\hline
$\hat{\sigma}_{B_c}$, \cite{ms}
              & 23.6    &   26.0  & 22.3    & 15.5    & 11.2    & 8.4     \\
\hline
\end{tabular}
\caption{Comparison of the total cross
         section in pb for $gg \rightarrow B_c b \bar{c}$ with the result of
         ref. [6].
         The input parameters are $m_b+m_c=6.3$ GeV,
         $m_b=3m_c$, $\alpha_s=0.2$ and $f_{B_c}=0.50$ GeV.
         The number in parenthesis shows the Monte Carlo uncertainty of
         the last digit.}
\label{mstab}
\end{center}
\end{table}

\begin{table}
\begin{center}
\begin{tabular}{|c|c|c|c|}
\hline
$\sqrt{\hat{s}}$ [GeV] & Ref.~\cite{cc} & fixed $\alpha_s$
                           & $\alpha_s(\epsilon\hat s)$  \\
\hline
\hline
        20             &  4.9 & 4.9 & 4.9  \\
\hline
        30             &  8.5 & 7.0 & 5.0  \\
\hline
        60             &  7.9 & 5.9 & 2.4  \\
\hline
\end{tabular}
\caption{Comparison of the total cross section
         $\hat\sigma (gg \rightarrow B_c b \bar c)$ (in pb)
         for fixed and running $\alpha_s$ with the result quoted in ref. [2].
         The input parameters are $m_b=4.9\,{\rm GeV},\ m_c=1.5\,{\rm GeV}$,
         and $f_{B_c}=0.48\,{\rm GeV}$. The normalization of $\alpha_s$
         is chosen such as to reproduce the  cross section at 20 GeV.
}
\label{cctab}
\end{center}
\end{table}

Finally, the hadronic cross sections
given in ref. \cite{s} seem to differ even by more than one order of
magnitude from our results. Again, since
the normalization of $\alpha_s(Q^2)$ is
not specified, we cannot make a more definite statement.

To conclude, all previous calculations \cite{cc,s,bls,ms} are in mutual
disagreement. Our calculation confirms the results on $\hat\sigma$ of
ref. \cite{bls} at
$\sqrt{\hat s}\le 100\,$GeV when  including the missing color factor
\cite{bls1} and taking $m_b=4.8\,$GeV. However, at $\sqrt{\hat s}=1\,$TeV,
a clear numerical discrepancy remains.

\section{Distributions}
%
Having discussed the total $B_c^{(*)}$ production cross sections, we now turn
to the  differential distributions in transverse momentum and rapidity.
Since they are very similar for $B_c$ and $B_c^*$, we
show them for the pseudoscalar meson only. Moreover, since the
sensitivity to the scale in $\alpha_s(Q^2)$ and $g(x,Q^2)$ is already
indicated in Fig.~2,
we shall present here results only for the scale $Q^2=p_T^2+M^2$.
Finally, we will
examine the fragmentation description
and determine the
region of validity of this approximation. In order to be consistent with the
complete calculation at the fixed order $\alpha_s^4$, we do
not take into account evolution effects in
the fragmentation function $D_{\bar b}(z)$.
These effects are studied in ref. \cite{PP2}.
Moreover, it
should be mentioned that $c$--quark production and fragmentation is
negligible.

Fig.~3  illustrates the main features of the subprocess
$gg\rightarrow B_c b\bar c$. We see that the fragmentation description
approaches the $p_T$--distributions
resulting from the complete calculation only in the tails of the
distributions. At $\sqrt{\hat s} \le 40\,$GeV a problem arises also when $p_T$
approaches $p_T^{max}$ because of mass ambiguities in the phase space
boundaries of the fragmentation approach. These effects increase as
$\hat s$ decreases and are largest for the prescription III of eq.
(\ref{prescription}).
On the other hand, if the physical phase space boundary is imposed on
 the fragmentation approach as in eq. (\ref{prescription}) case I,
the approximation slightly improves.
Comparing Fig.~3 with Fig.~5 of ref. \cite{klr}, one sees that in contrast
to $\gamma\gamma\rightarrow B_c b\bar c$ where heavy quark fragmentation
is completely subdominant even at large $p_T$, in
$gg\rightarrow B_c b\bar c$ it provides a good approximation at
large $\hat s$ and $p_T$.
Obviously, the presence of gluon self-couplings and
 color factors has a drastic influence on the relative importance
of the fragmentation and recombination.
%
\ \vspace{1cm}\\
\begin{minipage}[t]{7.8cm} {
\begin{center}
\hspace{-1.7cm}
\mbox{
\epsfysize=7.0cm
\epsffile[0 0 500 500]{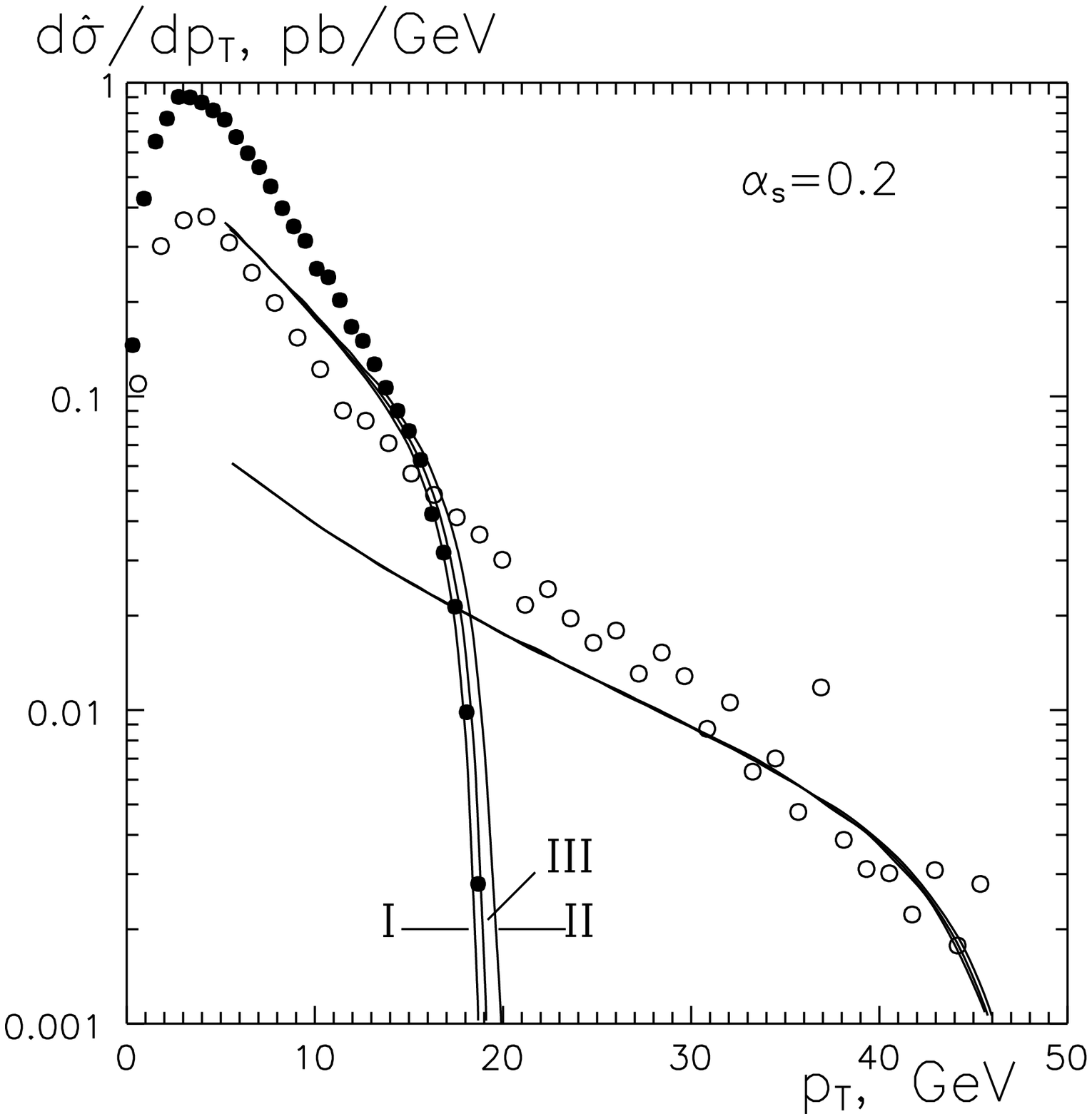}
}
\end{center}
\vspace*{-0.5cm}
\noindent
{\small\bf Fig.~3. }{\small\it
Transverse momentum distributions of the $B_c$ in
$gg\rightarrow B_cb\bar c$ at $\sqrt{\hat s} = 40$ and 100\,GeV:
complete $O(\alpha_s^4)$ calculation (circles) and fragmentation
approximation (solid curves). The labels I--III refer to the kinematics
specified in eq. (5).
}
}\end{minipage}
\hspace{0.5cm}
\begin{minipage}[t]{7.8cm} {
\begin{center}
\hspace{-1.7cm}
\mbox{
\epsfysize=7.0cm
\epsffile[0 0 500 500]{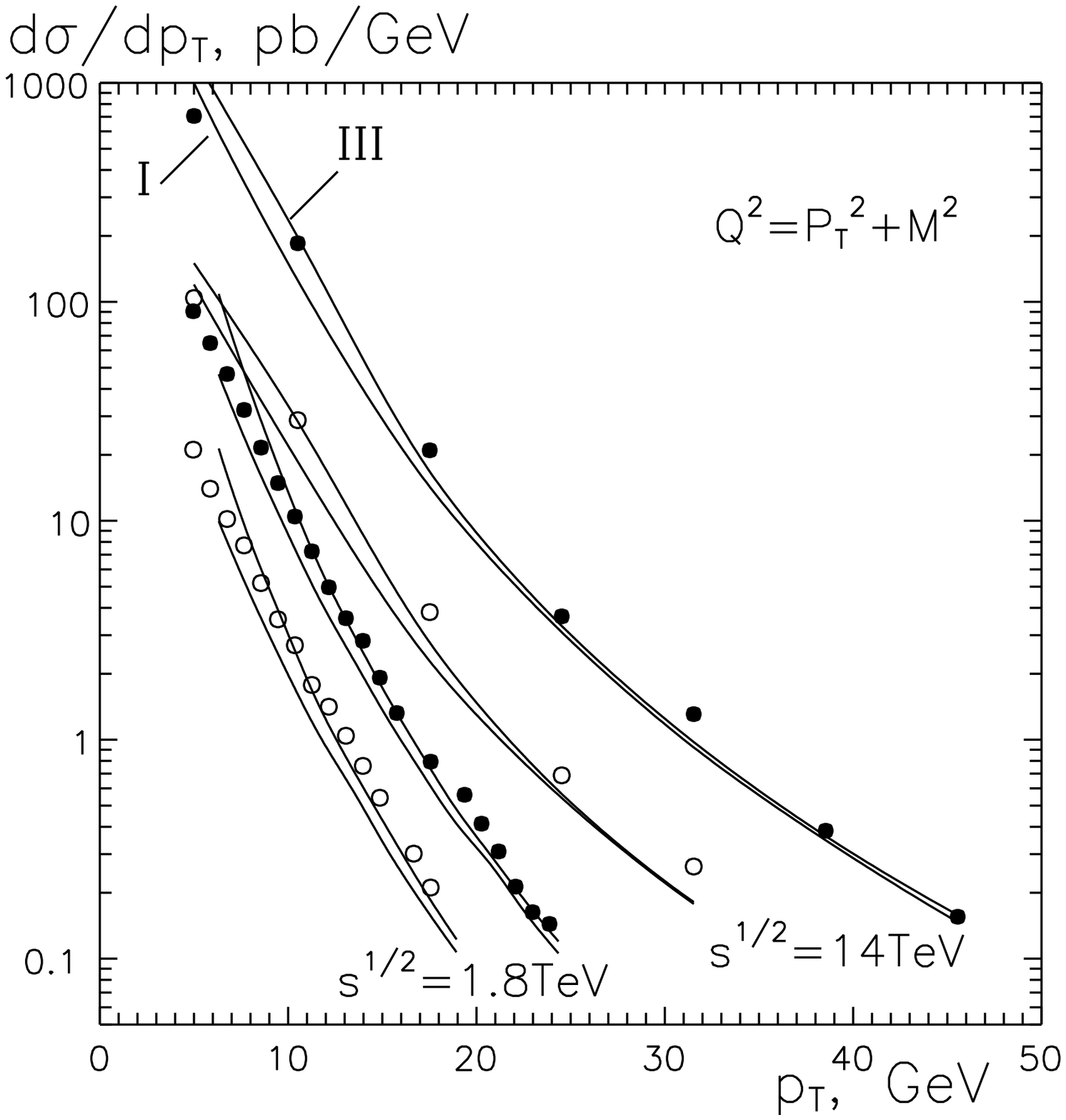}
}
\end{center}
\vspace*{-0.5cm}
\noindent
{\small\bf Fig.~4. }{\small\it
Transverse momentum distribution of the $B_c$ at Tevatron and LHC energies:
complete $O(\alpha_s^4)$ calculation (circles) and fragmentation
approximation (solid curves). The labels I and III refer to the kinematics
specified in eq. (5).
At each energy, results are shown without a rapidity cut (full circles) and for
 $|y|\leq 0.5$ (empty circles).
}
}\end{minipage}
\vspace*{0.5cm}

Distributions in $p_T$ for $pp$ (and $p\bar p$) collisions
are displayed in Fig.~4. Also shown is
the effect of a stringent cut in rapidity.
A few comments are in order. Most importantly, after convolution
the full calculation
 and the fragmentation description are in reasonable agreement at
$p_T\ge 10\,$GeV.
This can be understood from the properties of the unfolded
$p_T$--distributions illustrated in Fig.~3 and from
 the rise of the gluon density at small $x$ which favours contributions from
the smallest possible subenergies $\hat s$.
At $p_T < 10\,$GeV, the deviation grows indicating
that one is getting outside the range of validity of the
fragmentation picture.
Furthermore, the sensitivity to the kinematical prescription, eq.
(\ref{prescription}), decreases slowly with increasing $p_T$. Beyond the
$p_T$--region considered in Fig.~4, it can be safely ignored.
These observations are rather independent of the rapidity cut.
The rapidity distributions exhibit the usual plateau
region around $y=0$, as shown in Fig.~5.

Thus, in contradiction to ref. \cite{bls1}, we find that
the fragmentation description provides a good approximation for
$p_T\ge 10\,$GeV.
The deviation from the complete $O(\alpha_s^4)$ results is much smaller than
 the uncertainties in  the scale $Q^2$, in the decay constant
 $f_{B_c^{(*)}}$, and in the effective values of $m_b$ and $m_c$.
\ \vspace{1cm}\\
\begin{minipage}[t]{7.8cm} {
\begin{center}
\hspace{-1.7cm}
\mbox{
\epsfysize=7.0cm
\epsffile[0 0 500 500]{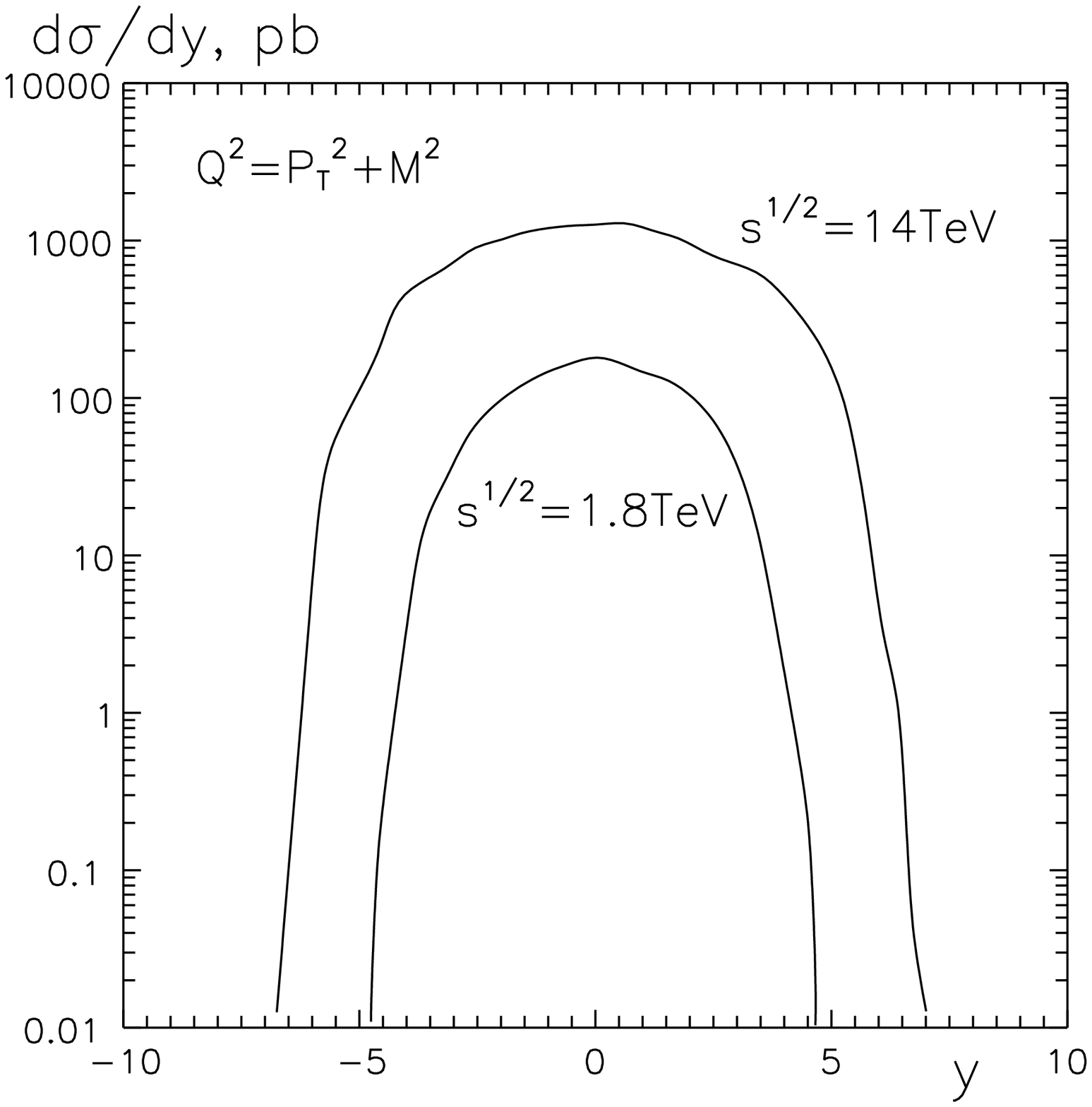}
}
\end{center}
\vspace*{-0.5cm}
\noindent
{\small\bf Fig.~5. }{\small\it
The rapidity distributions of the $B_c$ mesons at Tevatron and LHC energies.
}
}\end{minipage}
\hspace{0.5cm}
\begin{minipage}[t]{7.8cm} {
\begin{center}
\hspace{-1.7cm}
\mbox{
\epsfysize=7.0cm
\epsffile[0 0 500 500]{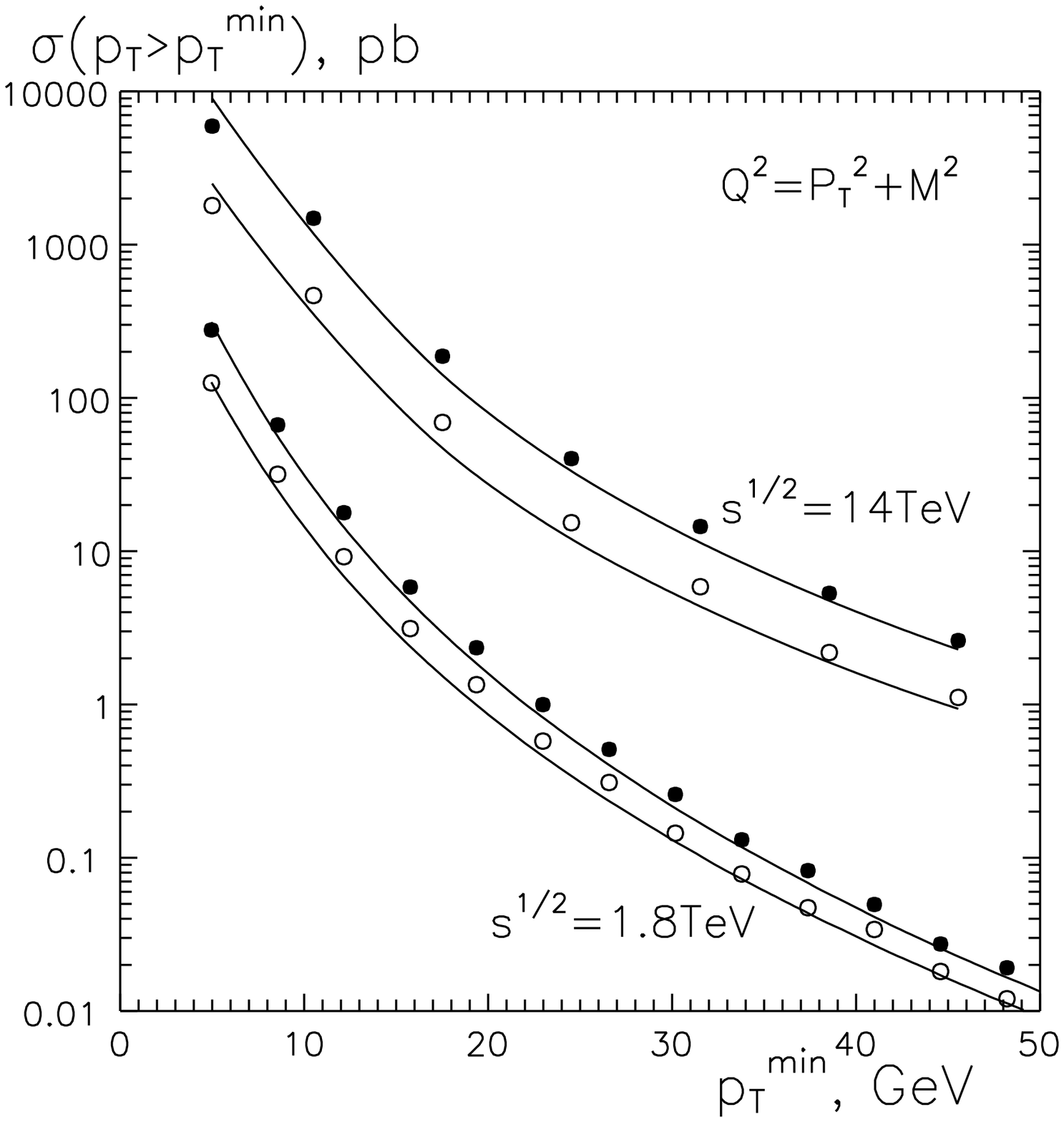}
}
\end{center}
\vspace*{-0.5cm}
\noindent
{\small\bf Fig.~6. }{\small\it
Integrated production cross sections for the $B_c$ as functions of the
minimum $p_T$ cut:
complete $O(\alpha_s^4)$ calculation (circles) and fragmentation approximation
using eq. (5) II (solid curves).
For each energy, results are shown without a rapidity cut (full circles)
and for $|y|\leq 1$ (empty circles).
}
}\end{minipage}
\section{Summary and conclusions}
We have presented new results on  $B_c$ meson production in $pp$
and $p\bar p$ collisions derived from the $O(\alpha_s^4)$ subprocess
$gg\rightarrow B_c^{(*)}b\bar c$ using
 nonrelativistic bound state approximations. We have performed
two independent calculations
employing two different methods. We therefore believe that our results which
deviate from all  previous calculations \cite{cc,s,bls,ms,bls1} are correct.

In addition, we have determined the range of
validity of the fragmentation approximation with
 the heavy quark fragmentation
function $D_{\bar b}(z)$ as calculated in perturbation theory \cite{fragee}.
For $p_T\ge 10\,$GeV, we find good
quantitative agreement between the fragmentation description and the
complete calculation. At $p_T = 5 - 10\,$GeV, mass ambiguities
 in the fragmentation kinematics
lead to effects up to a factor of two. Below 5\,GeV, the fragmentation
approximation brakes down.

In order to evaluate the observability of $B_c$ mesons at the Tevatron and LHC,
it is useful to integrate the $p_T$--distributions of Fig.~4 over
$p_T\ge p_T^{min}$. The resulting cross sections are plotted in Fig.~6.
This estimate indicates a  production of the order of $10^4\ B_c$
mesons with $p_T^{min}= 10\,$GeV at the Tevatron, assuming an integrated
luminosity of $100\,{\rm pb}^{-1}$.
Contributions from the production and decay of $B_c^*$ mesons and heavier
states are not yet taken into account in this number (see e.g. ref.
\cite{PP2}). This rate may be sufficient for a first observation
in the decay channels $B_c\rightarrow J/\psi X$, for
which a branching ratio of the order of 10\% is predicted \cite{decay}.
Finally, at the LHC for 100${\rm fb}^{-1}$ one can expect $10^7$
 direct $B_c$
mesons  at $p_T^{min}= 20\,$GeV which should make $B_c$ physics a very
interesting topic.
%

\end{document}